\documentclass{article}

\usepackage{PRIMEarxiv}

\usepackage[utf8]{inputenc} 
\usepackage[T1]{fontenc}    
\usepackage{hyperref}       
\usepackage{url}            
\usepackage{booktabs}       
\usepackage{amsfonts}       
\usepackage{nicefrac}       
\usepackage{microtype}      
\usepackage{lipsum}
\usepackage{fancyhdr}       
\usepackage{graphicx}       
\usepackage{subfigure}
\graphicspath{{media/}}     

\title{Band Gap Tuning of DC Reactively Sputtered ZnON Thin Films
\thanks{\textit{\underline{Citation}}: 
\textbf{Authors. Title. Pages.... DOI:000000/11111.}} 
}

\author{
  Kiran Jose, \\
  National Institute of Technology Calicut \\
  Kozhikode, Kerala, India \\
  \texttt{kiranpatteriyil@gmail.com} \\
   \And
  JG Anjana \\
  National Institute of Technology Calicut \\
  Kozhikode, Kerala, India \\
  \texttt{anjana\textunderscore p180026ec@nitc.ac.in} \\
  \AND
  Venu Anand \\
  National Institute of Technology Calicut \\
  Kozhikode, Kerala, India \\
  \texttt{venuanand@nitc.ac.in} \\
  \AND
  Aswathi R Nair \\
  National Institute of Technology Calicut \\
  Kozhikode, Kerala, India \\
  \texttt{aswathirnair@nitc.ac.in} \\
}

\begin{document}
\maketitle

\begin{abstract}
Zinc oxynitride (ZnO$_x$N$_y$) has recently emerged as a highly promising band gap-tunable semiconductor material for optoelectronic applications. In this study, a novel DC reactive sputtering protocol was developed to fabricate ZnO$_x$N$_y$ films with varying elemental concentrations, by precisely controlling the working pressure. The band gap was rigorously analyzed using UV-Visible spectroscopy, which was complemented by EDAX spectroscopy to determine the variations in the elemental composition. The correlation between the microstructure and band gap was investigated through the application of AFM, XRD, and Raman spectroscopy, while the Urbach theorem was used to evaluate the defect states. This study revealed the existence of intermediate structures formed during the tuning of the band gap, which can have important implications for future research aimed at developing heterostructures and 2D superlattices for photonics applications.
\end{abstract}


\section{Introduction}
\label{section1}
The optoelectronics industry is constantly seeking materials with excellent electrical and optical properties to meet the demands of modern computational algorithms. Amorphous semiconductors are a versatile class of materials that have found wide-spread applications in the field of electronics and flexible electronics, including in the areas of thin-film transistors (TFTs)\cite{bhalerao2022flexible}, LCD displays\cite{ma2019robust}, solar cells\cite{yu2016metal}, sensors for gas\cite{ahmadipour2022detection}, humidity, temperature, and pressure detection\cite{nunes2019metal}, memory devices, neuromorphic devices\cite{hong2018oxide}, photodetectors, photovoltaics\cite{chen2020recent}, and spintronics\cite{c2014oxides}. However, low electron mobility, a wide band gap, and persistent photoconductivity (PPC)\cite{nathan2013amorphous} of amorphous semiconductors, such as oxides, limit the performance and efficiency of devices such as phototransistors and sensors, making it difficult to meet the demands of modern technology. Researchers have turned to alternative materials like zinc oxynitride (ZnO$_x$N$_y$) which is a band gap tunable semiconductor material with high electron mobility of up to 100V/s\cite{jeong2016investigation} for improved performance and efficiency in devices like phototransistors\cite{ryu2012high}. Additionally, ZnO$_x$N$_y$ has a low persistent photoconductivity (PPC)\cite{lee2015localized}\cite{jang2020influence}, which means it does not store electrical charges for a long time after the light is turned off; this improves the operational stability of the device. These properties make ZnO$_x$N$_y$ a prospective substitute for optoelectronics and photonics applications.

One of the most promising forms of this material is a thin film, which has been the subject of a number of recent research studies\cite{park2016study}\cite{anjana2022surface}\cite{kim2022nonvolatile}. These studies have investigated the properties and potential applications of ZnO$_x$N$_y$ thin films, including their optical and electronic properties[16-20], as well as their potential use in devices such as light-emitting diodes (LEDs)\cite{park2022highly}, solar cells, and sensors\cite{lee2014nanocrystalline}. Recent research has also focused on developing methods for synthesising high-quality ZnO$_x$N$_y$ thin films, including chemical vapour deposition (CVD)\cite{akinwunmi2022characterisation} and pulsed laser deposition (PLD) techniques\cite{ayouchi2014zinc}.

Sputtering is a well-established technique for depositing thin films of various materials, including ZnO$_x$N$_y$\cite{park2016study}. In reactive sputtering, a target material is bombarded with energetic ions in the presence of various reactive gas ions, which causes the target atoms to be ejected and deposited onto a substrate to form a thin film. One of the key advantages of this technique is the ability to control the stoichiometry and composition of the deposited film by adjusting the sputtering conditions and the composition of the target material\cite{berg2005fundamental}. Studies have shown that it is possible to deposit ZnO$_x$N$_y$ thin films with varying nitrogen content, which can affect the film's electrical and optical properties, making it a powerful technique for depositing ZnO$_x$N$_y$ thin films for various applications\cite{park2016study}\cite{park2022highly}\cite{lee2014nanocrystalline}.

However, there are challenges associated with reactively sputtering ZnO$_x$N$_y$ thin films. These include a decrease in sputtering yield due to the reacting gas, the formation of a layer of deposited film in the target (target poisoning), and the difference in reactivity of gases when more than one gas is used. Deposition of ZnO$_x$N$_y$ film involves three gases, Argon the sputtering gas, Nitrogen with relatively low reactivity and highly reacting Oxygen, which leads to further complexity. Thus, the presence of residual oxygen can be a major challenge in preparing ZnO$_x$N$_y$ thin films and can even call into question the repetition of the experiment.

The ability to tune the band gap is a significant aspect of ZnO$_x$N$_y$ that draws attention of researchers for its potential use in optoelectronic devices, such as photosensors\cite{jang2015study} and high-mobility transistors\cite{ye2009high}. The careful selection of the band gap region is of utmost importance, as it has a direct correlation with the morphology, crystal structure, and defect states within the semiconductor, which can greatly impact the device's performance. Hence, the regulation of the band gap and the comprehension of the related microstructures of ZnO$_x$N$_y$ thin films are imperative for successful utilization in optoelectronic devices.

This work endeavours to address the challenges encountered during the sputtering process of ZnO$_x$N$_y$ thin films, with the objective of finding the optimal conditions for tuning the band gap of the deposited film. The protocol involves adjusting the pressure of the sputtering process by introducing various partial pressures of nitrogen, while keeping the other sputtering parameters constant. Even though a direct current (DC) power source was utilized in this work, the consideration of a radio frequency (RF) power source is also being explored as a potential solution to mitigate target poisoning, which is a prevalent issue in sputtering. This is achieved through the prevention of ion buildup on the target material, facilitated by the alternating polarity of the RF power source. 

Once the deposition conditions have been optimized for different band gaps, the study aims to delve deeper into the characterization of the film, examining the evolution of its structural phases, morphology, and defect states as the band gap changes. This comprehensive analysis will be carried out using various state-of-the-art characterization techniques such as Atomic Force Microscopy(AFM), X-ray Diffraction Analysis (XRD), Raman Spectroscopy, and UV-Visible Spectroscopy, to provide a comprehensive understanding of the properties of the ZnO$_x$N$_y$ thin film.

\section{Experimental Details}
\label{section2}

\subsection{DC Reactive Sputtering}
\label{subsection1}

In this study, ZnO$_x$N$_y$ thin films were prepared using a reactive sputtering technique. A zinc target was sputtered using argon in the presence of both oxygen and nitrogen gases with a DC power source. The major sputtering parameters, including the target material and size, DC power, base pressure, sputtering power, the flow rate of different gases etc., are listed in Table \ref{table1}. The dc power for all other depositions except for studying the effect of sputtering power was 35 W. The substrates used for characterization were silicon for AFM, Energy Dispersive X-ray Analysis (EDAX), XRD, Raman spectroscopy and Corning Plain Microscope Glass Slide for UV-visible spectroscopy.

\begin{table}[!t]
\caption{Sputtering Parameters. }
\label{table1}
\centering

\begin{tabular}{|l |  l|}

\hline
Target & 3" Zinc\\
\hline
Base Pressure & 4.2 X 10$^{-6}$ mbar\\
\hline
Argon Flow Rate&4 to 7 sccm\\
\hline
Nitrogen Flow Rate & 10 to 25 sccm\\
\hline
Power & 35 W\\
\hline
Substrate Rotation & 20 rpm\\
\hline
Substrate Temperature & Room Temperature\\
\hline
\end{tabular}
\end{table}

The substrates were cleaned by sonicating for 15 minutes in acetone, followed by isopropyl alcohol and doubly deionized water. The target was also cleaned by pre-sputtering using an RF plasma before every sputtering, and the sputtering time was kept to a minimum to reduce the target poisoning.

During the initial phase of the study, the strategy for optimizing the band gap of ZnO$_x$N$_y$ thin film deposition was carefully evaluated. This was achieved by maintaining a constant flow rate of argon at 5 sccm while adjusting the flow rate of nitrogen from 15 to 24 sccm. It is noteworthy that during the deposition process, the chamber was not supplemented with any additional oxygen through the use of mass flow controllers. Instead, the residual oxygen that remained within the chamber after attaining the base pressure was utilized.

\subsection{Tauc Plot}
\label{subsection2}

The band gap of ZnO$_x$N$_y$ thin films is determined from UV-visible spectroscopy by employing Tauc equation\cite{tauc2012amorphous},

\begin{equation} \label{e1}
\begin{array}{ll}
\alpha h \nu = A(h \nu-E_g)^n 
\end{array}
\end{equation}

Where $\alpha$ is the absorption coefficient, $h\nu$ is the photon energy, A is a constant, E$_g$ is the band gap energy, and $n$ is a material constant that is typically taken as 2 for direct band gap semiconductors and 1/2 for indirect band gap semiconductors for allowed transitions. Here, we plot ($\alpha$h$\nu$)$^2$ with h$\nu$\cite{gomez2018identification}, and by analyzing the linear portion of the Tauc plot, the direct band gap energy (Eg) is determined by extrapolating to the energy axis.

\subsection{Determining Thickness}
\label{subsection3}

The deposition rate of the thin film was found by dividing the thickness of the film by the sputtering time. AFM was used to find the thickness of the film. For this, a portion of the silicon wafer was masked, so that no film would be formed in the masked area. By measuring the film height at this step, we estimated the thickness of the film. The roughness parameter of the film was also obtained from the AFM measurement.

\subsection{Urbach Energy}
\label{subsection4}

The Urbach tail energy can be calculated by analyzing the absorption spectra obtained from UV-visible spectroscopy using Urbach's rule. The Urbach rule describes the exponential decay of the absorption coefficient as a function of energy and is given by\cite{urbach1953long}

\begin{equation} \label{e2}
\begin{array}{ll}
\alpha(E) = \alpha_0 exp(E/E_u) 
\end{array}
\end{equation}

where E is the energy and $E_u$ is the Urbach tail energy. To calculate the Urbach tail energy, the natural logarithm of the absorption coefficient, ln($\alpha$), is plotted as a function of energy, E. The slope of this plot, just before the band gap energy, is given by -1/$E_u$. So the Urbach tail energy, $E_u$, can be calculated by taking the reciprocal of the slope\cite{isik2023growth}.

\section{Results and Discussion}
\label{section3}

\subsection{Band Gap Engineering}
\label{subsection5}

\subsubsection{Band Gap Tuning}
\label{subsubsection1}

\begin{figure*}
\centering
\subfigure{
   \includegraphics[scale =0.27]{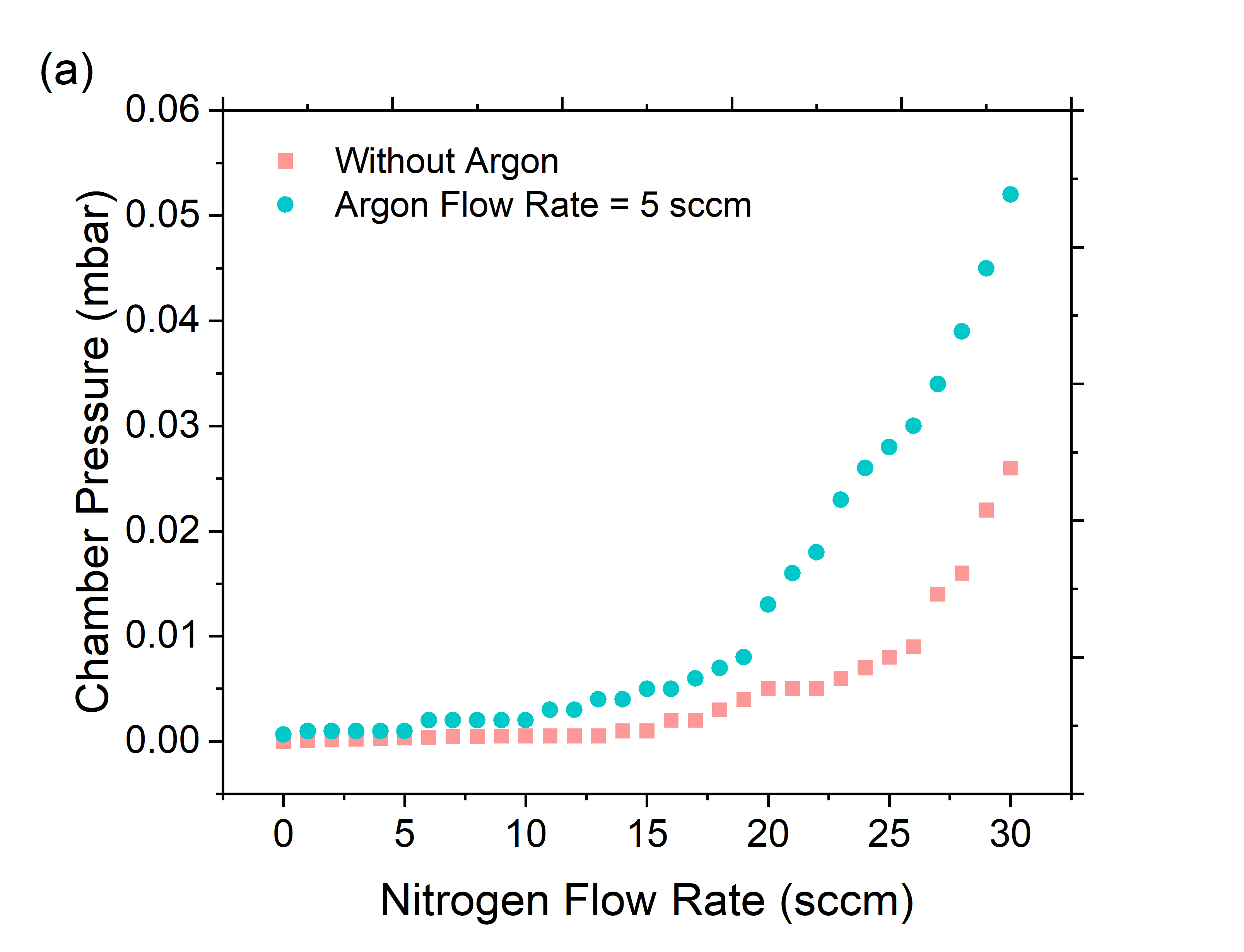}
   \label{fig1a}
 }
 \subfigure{
   \includegraphics[scale =0.27]{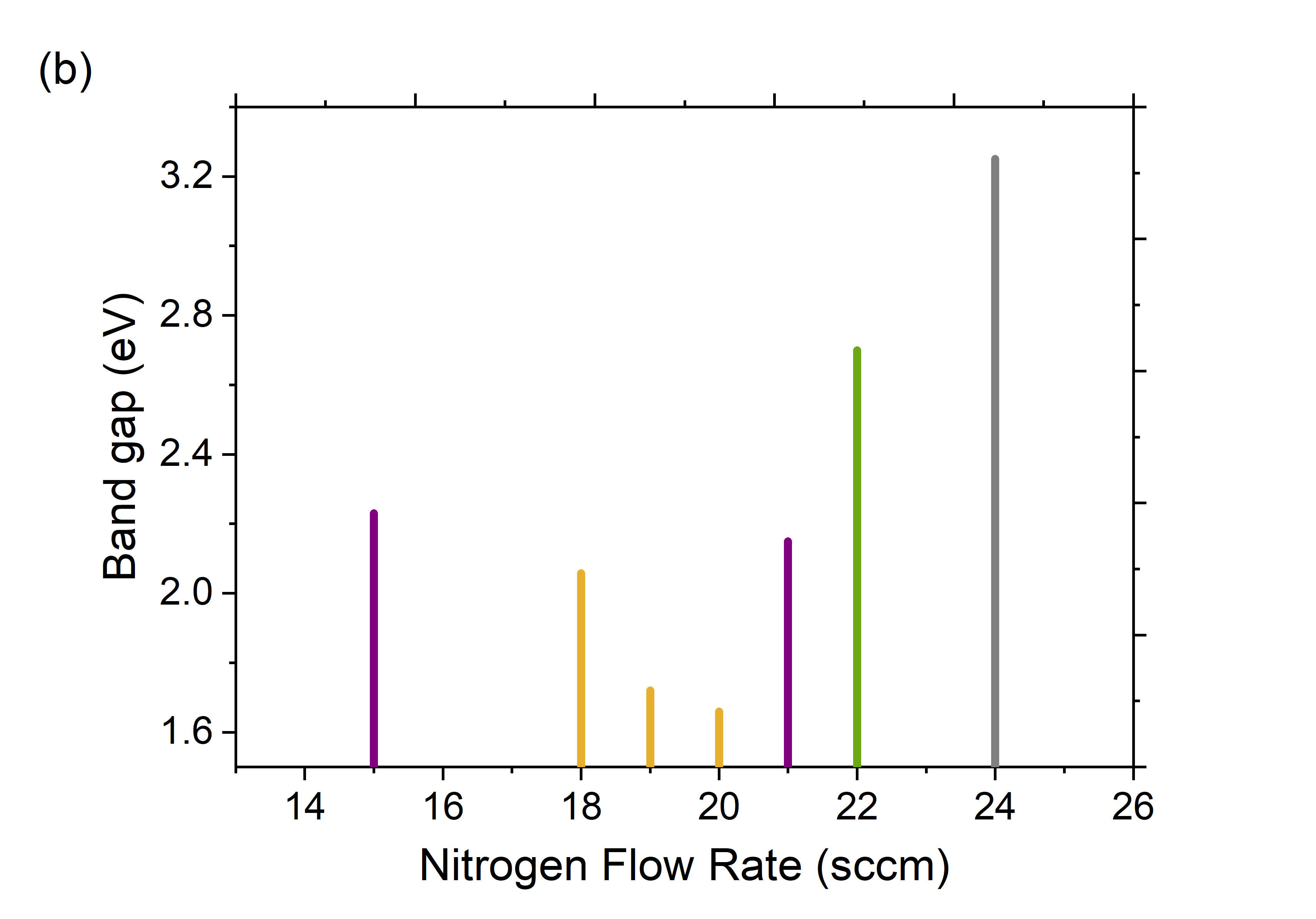}
   \label{fig1b}
 }
\subfigure{
   \includegraphics[scale =0.27]{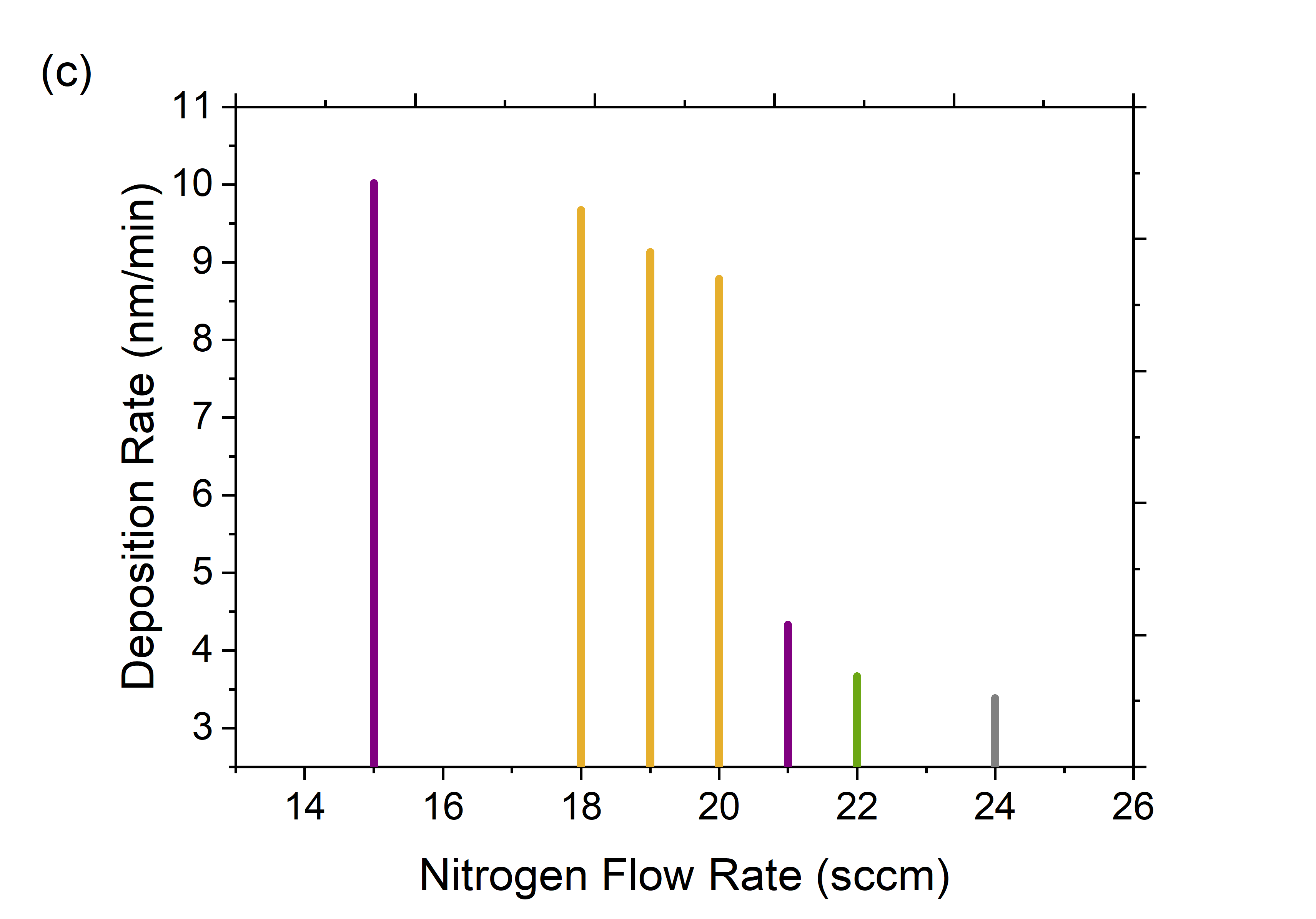}
   \label{fig1c}
 }
 \subfigure{
   \includegraphics[scale =0.27]{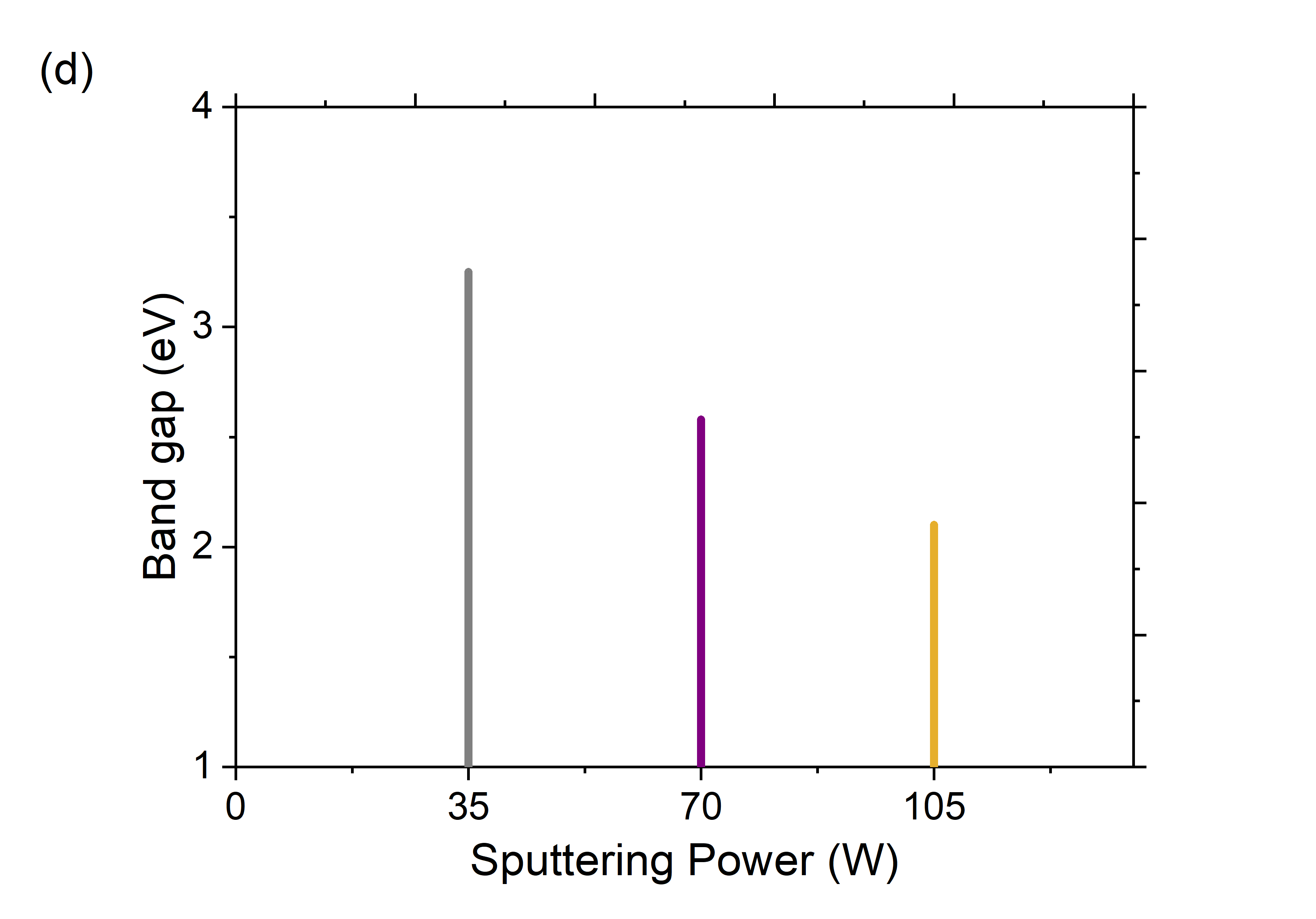}
   \label{fig1d}
 }
 
\label{1}
\caption{Variations of Chamber Pressure, Band Gap, Deposition Rate, and Band Gap with Nitrogen Flow Rate and Sputtering Power. (a) Chamber pressure vs nitrogen flow rate after evacuation to 4.4 x 10$^{-6}$ mbar (pink) and with 5 sccm argon flow rate after evacuation (cyan), (b) Band gap vs nitrogen flow rate, (c) Deposition rate vs nitrogen flow rate, (d) Band vs sputtering power with constant other sputtering parameters. The color pattern used in (b), (c), and (d) corresponds to different structural phases which will be discussed later. For the denotation of different colors, please refer to the caption of Figure 3.}
\end{figure*}

The way to tune the band gap of ZnO$_x$N$_y$ is by varying the composition of the thin film. A combination of a higher atomic concentration of nitrogen and lower concentration of oxygen will give a low band gap and vice versa\cite{kaczmarski2016fabrication}. The detailed study of the relation between composition and band gap can be seen in section \ref{subsubsection2}. Previous research has focused on altering nitrogen-to-oxygen ratios\cite{ryu2012high} at high sputtering power\cite{ievtushenko2019effect} and working pressures\cite{reinhardt2016electron} to produce varying compositional stoichiometries. In the present study, we suggest a novel method for accomplishing this with low working pressure, low power, and utilisation of residual oxygen without introducing further oxygen to the process. 

In earlier studies, the ZnO$_x$N$_y$ thin films were developed by alternating a 100 to 200 SCCM flow rate of nitrogen with 1 to 10 SCCM of oxygen\cite{jang2020influence} \cite{reinhardt2016electron} to balance the disparity in reactivity between the two gases. However, our approach aims to achieve the same result by varying the working pressure inside the chamber. The challenge of incorporating nitrogen in the film is due to the high ionization energy of the nitrogen molecule compared to the oxygen molecule. Since nitrogen atoms are triple-bonded to form a molecule, high energy is required to dissociate this bond\cite{zhang2007characterization}. As a result, most of the energy will be used for the ionization of the nitrogen molecule, and the remaining will only be reflected in the kinetic energy. 

In order to achieve varying incorporation of nitrogen in the film, we varied the flow rate of nitrogen gas while keeping other parameters constant. The mass flow rate of Argon was kept at 5sccm, and the flow rate of nitrogen was varied from 15 sccm to 24 sccm, with a sputtering power of 35W. The gate valve was kept completely open during all the experiments. The working pressure was monitored and recorded during the experiment; the results are shown in Fig.\ref{fig1a}. The graph shows an exponential increase in working pressure as the flow rate of nitrogen increases. 

It is seen that (Fig.\ref{fig1b}) the band gap of the film initially decreases and reaches a minimum of 1.6 eV at a flow rate of 20 sccm of nitrogen, after which it drastically increases. This suggests that the incorporation of nitrogen increases as the flow rate of nitrogen increases, reaching a maximum at this point and then falls. The initial trend of increased nitrogen incorporation is a result of the rise in the flow rate of nitrogen, which in turn elevates the partial pressure of nitrogen within the chamber. This finds a \emph{optimum point} at a certain partial pressure where there is enough amount of nitrogen. However, the fall in the incorporation is due to increased working pressure. After reaching the \emph{optimum point}, an increase in working pressure leads to a decrease in the number of nitrogen ions that reach the substrate. As discussed earlier, nitrogen ions will have the least kinetic energy among other ions because of the high ionization energy. Despite the high concentration of nitrogen within the chamber, the nitrogen ions may not possess sufficient energy to reach the substrate due to a large number of collisions caused by increased working pressure. This, in turn, leads to a decrease in nitrogen incorporation as the nitrogen flow rate continues to increase.

The trend in the deposition rate of the film as a function of the nitrogen flow rate, as depicted in Fig.\ref{fig1c}, displays a persistent decline as the concentration of gases within the chamber increases. This initial decline in the deposition rate, which is less pronounced, can be attributed to the conventional phenomenon of reactive sputtering, in which the sputtering rate decreases with increasing amounts of reactive gas. However, after a nitrogen flow rate of 20 sccm, a significant decline in the deposition rate can be observed. This decline is the result of the increased working pressure, as previously discussed, which results in an increase in the number of collisions and a reduction in the probability of the sputtering ions reaching the substrate.

The tuning of the band gap in this study was achieved by controlling the mean free path of ions, specifically nitrogen ions, within the sputtering chamber. The mean free path refers to the average distance travelled by a particle between two consecutive collisions in a gas or solid. In a gaseous medium, it represents the average distance a molecule can travel without encountering another molecule. Before a working pressure of 0.013 mbar (achieved through the application of a nitrogen flow rate of 20 sccm and an argon flow rate of 5 sccm after attaining a base pressure of 4.2 x 10$^{-6}$ mbar), the ions possessed sufficient path to traverse without collision. At this stage, the mere increase in the amount of reactive gases resulted in a corresponding increase in the incorporation of the element. However, as the nitrogen flow rate was increased to 21 sccm, the working pressure increased to 0.016 mbar, leading to a greater number of collisions and a reduction in the mean free path of the ions. This decrease affected all reactive ions, with a significant drop in deposition rate observed in this region. The impact was particularly pronounced for nitrogen, as a result of its high ionization energy and corresponding lower kinetic energy, leading to a greatly reduced mean free path. As the flow rates were increased further, the working pressure increased nearly exponentially, leading to an exponential decrease in nitrogen incorporation.

The variation of the band gap with respect to the power is presented in Fig.\ref{fig1d}, reinforcing our proposed theory. As previously noted, the band gap is affected by the mean free path of nitrogen ions in the sputtering chamber. The lower kinetic energy of the nitrogen ions makes it less likely for them to reach the substrate when the working pressure is high, resulting in reduced incorporation of nitrogen. However, by increasing the power, the kinetic energy of the nitrogen ions can also be increased, leading to a higher mean free path and a greater probability of nitrogen ions reaching the substrate.

It was observed that when the flow rate of argon was kept constant at 5 sccm and the nitrogen flow rate was increased to 24 sccm, the band gap was measured to be 3.25 eV at a sputtering power of 35W. This suggests low nitrogen incorporation in the sample. However, increasing the sputtering power to 70W and 105W under the same conditions led to a reduction of the band gap to 2.58 eV and 2.2 eV, respectively (as depicted in Fig.\ref{fig1d}). This shows that as the power increases, so does the kinetic energy of the nitrogen ions and their mean free path, leading to a higher incorporation of nitrogen into the ZnO$_x$N$_y$ thin films. These results support our theory about controlling the mean free path of nitrogen ions for band gap tuning and suggest that adjusting the power can also be used as an alternative method, in addition to controlling working pressure.

\subsubsection{Band Gap and Elemental Composition}
\label{subsubsection2}

\begin{figure*}
\centering
\subfigure{
   \includegraphics[width=8cm]{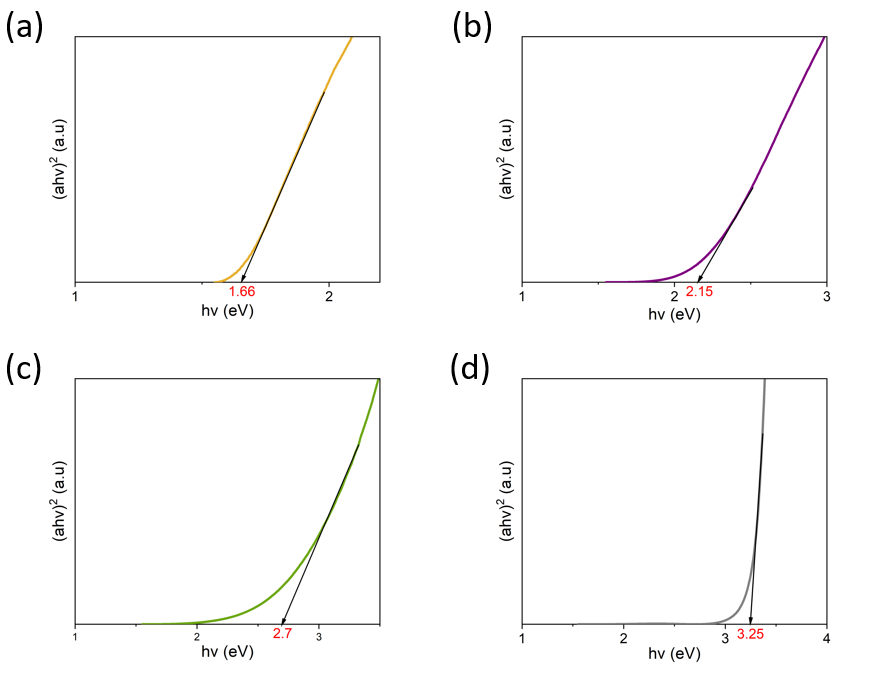}
   \label{fig2a}
 }
 \subfigure{
   \includegraphics[width=9.5cm]{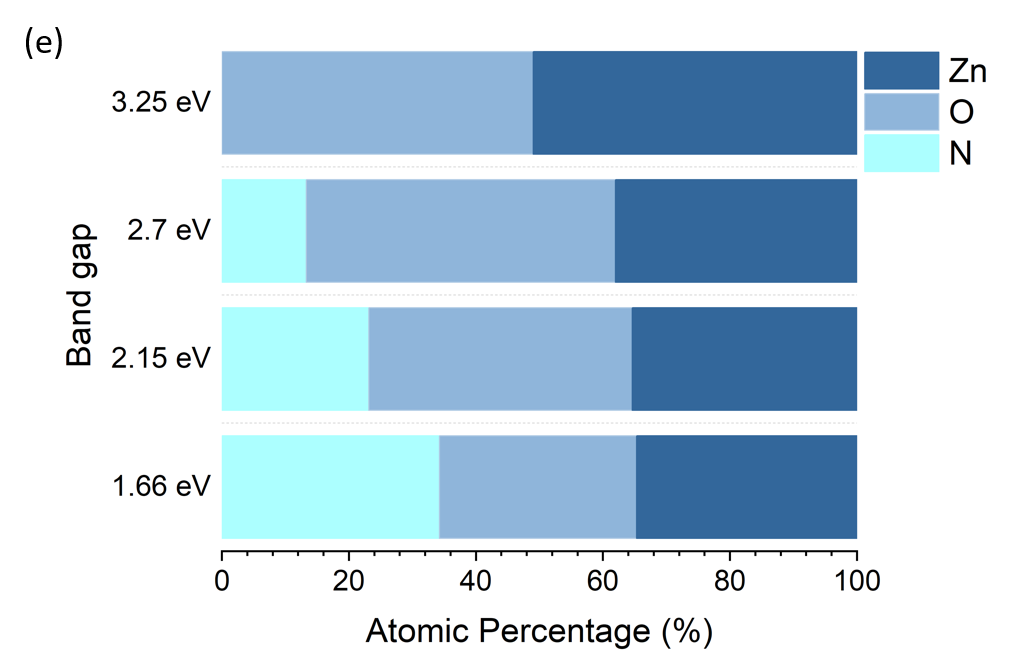}
   \label{fig2b}
 }

\label{2}
\caption{Variation of Band Gap and Atomic Composition in ZnO$_x$N$_y$ Thin Films. (a)-(d) Tauc Plot showing the variation of band gaps obtained from UV-Visible spectra for 1.66 eV, 2.15 eV, 2.7 eV, and 3.25 eV, respectively. (e) Atomic composition of the films with these band gaps obtained from EDAX spectroscopy.}
\end{figure*}

 The band gap of the ZnO$_x$N$_y$ thin films was characterized using a Tauc plot, derived from the transmittance spectra (as shown in Fig.\ref{fig2a}). Further insight into the variation in the material's elemental composition with band gap was obtained through EDAX spectroscopy, as shown in Fig.\ref{fig2b}. Our observations reveal a clear correlation between the decreasing band gap of the material and the increasing concentration of nitrogen within the film.

 ZnO$_x$N$_y$ is an alloy derived form from parental compounds Zinc nitride (Zn$_3$N$_2$) and Zinc oxide (ZnO). The band gap of pure Zn$_3$N$_2$ is reported to be 1.1 eV, while the band gap of pure ZnO is 3.25 eV\cite{lee2015localized}. One of the main reasons for this difference in band gap is the relative energy levels of the N 2p orbitals and O 2p orbitals\cite{kaczmarski2016fabrication}. In Zn$_3$N$_2$, the N 2p orbitals, which are observed above the O 2p orbitals, act as the valence band while the conduction band is dominated by the spherical Zn 4s orbital, similar to the case of ZnO\cite{xian2016structural}\cite{reinhardt2016electron}. This results in a lower band gap of Zn$_3$N$_2$ compared to ZnO. And when it comes to the case of ZnO$_x$N$_y$, as the atomic concentration of nitrogen increases, more electrons occupy the 2P level of nitrogen and start acting as the valence band, which in turn reduces the band gap\cite{kim2013anion}\cite{jang2020influence}.

\subsection{Structural Transition}
\label{subsection6}

\begin{figure*}[t!]
\centering
\subfigure{
   \includegraphics[width=8cm]{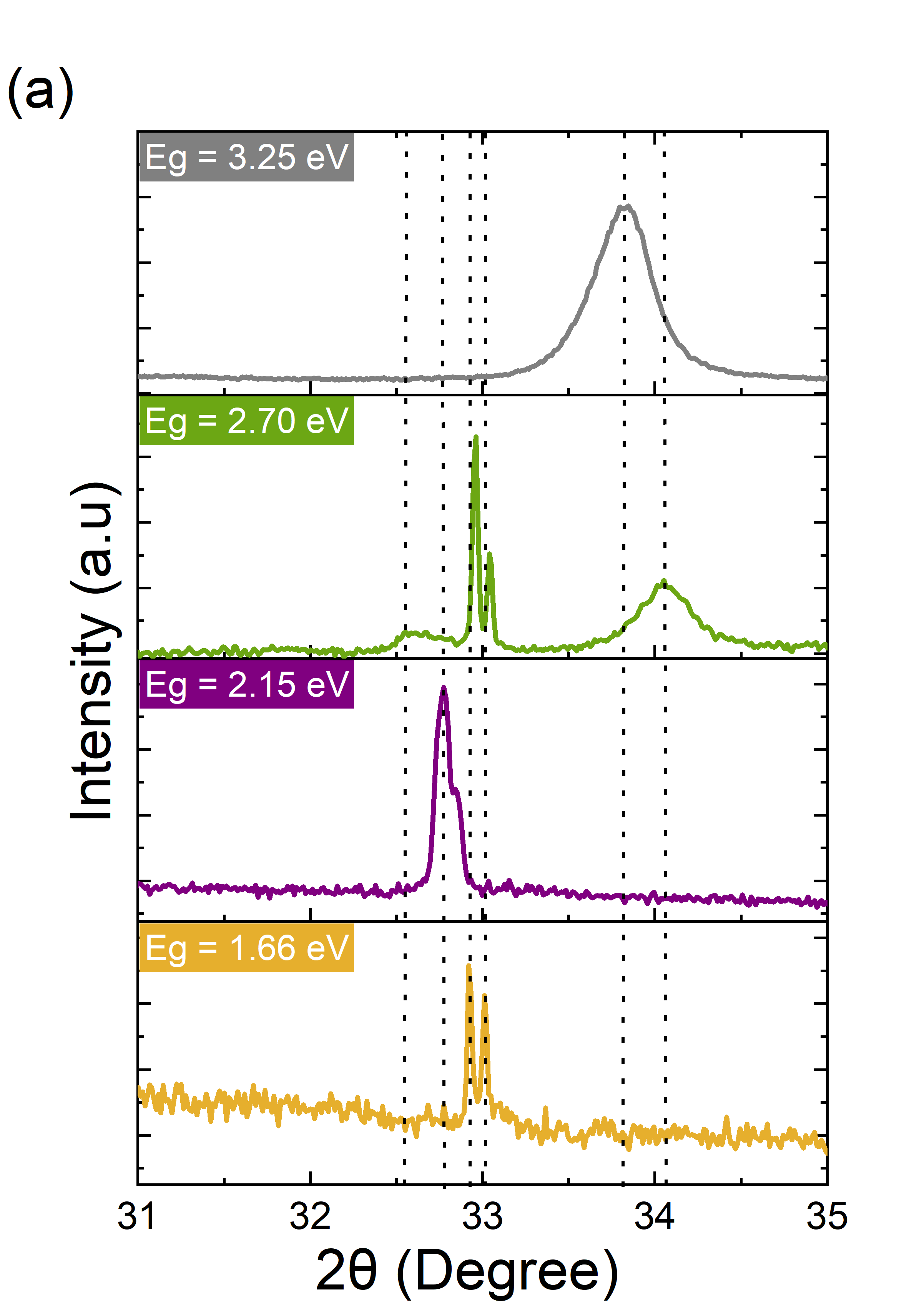}
   \label{fig3a}
 }
 \subfigure{
   \includegraphics[width=8cm]{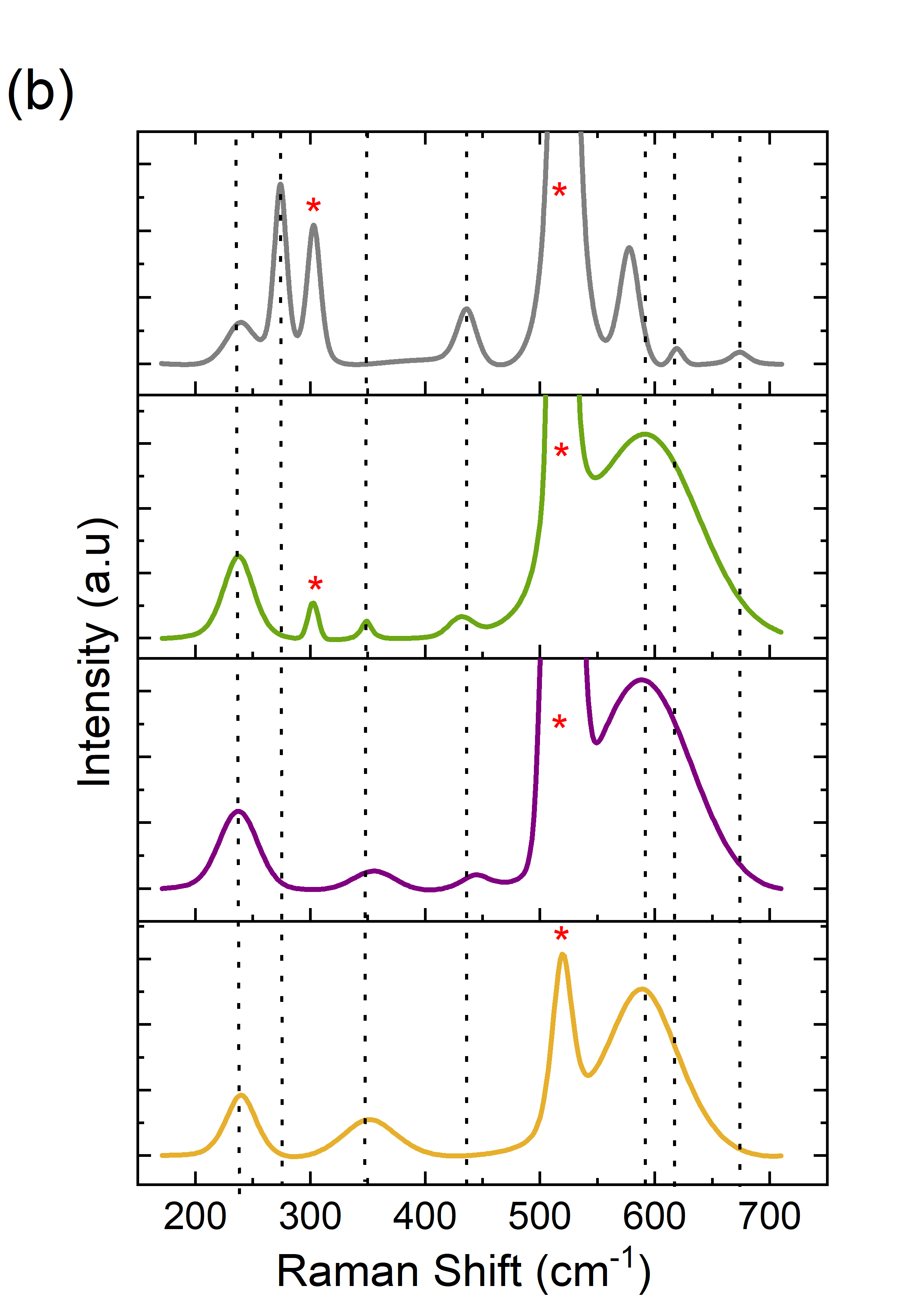}
   \label{fig3b}
 }

\label{3}
\caption{XRD and Raman Patterns for Thin Films with Different Band Gaps. (a) XRD and (b) Deconvoluted Raman patterns show different structures corresponding to different band gaps. Yellow colour represents pure alloy structure; purple colour represents wurtzite phase formation in the alloy platform; green colour represents the transition from alloy to wurtzite with a prominent wurtzite phase; grey colour represents distorted wurtzite structure without any alloy phase. This colour pattern is consistent throughout the article. Note: The peaks marked with a red star ($*$) are related to the substrate.
}
\end{figure*}

 In this section, we aim to gain a comprehensive understanding of the structural evolution in ZnO$_x$N$_y$ thin films as a function of the band gap. Previous studies have established a transition from the cubic structure to a hexagonal wurtzite structure as the band gap increases. However, there remains a significant gap in our understanding of the intermediate phases involved in this structural evolution. To address this, we employ a combined approach of XRD and Raman spectroscopy to provide a thorough characterization of the structural changes. XRD measurements were performed using a Malvern Panalytical system equipped with Cu K$\alpha$ radiation, while Raman spectra were collected after exciting the molecules with a 532 nm wavelength laser on a Labram HR Evolution confocal Raman microscope. The integration of these techniques provides a comprehensive evaluation of the structural properties of the ZnO$_x$N$_y$ thin films and the transitions involved in their formation.

 The XRD patterns of ZnO$_x$N$_y$ thin films often showcase distinctive peaks within the 2$\theta$ range of 31$^0$ to 35$^0$, offering a comprehensive understanding of the structural properties of the films. A prominent peak reveals the hexagonal wurtzite structure of ZnO at approximately 34.442$^0$ 2$\theta$ (JCPDS file, card no. 01-074-0534), while a peak indicates the cubic structure of Zn$_3$N$_2$ at approximately 31.3$^0$ 2$\theta$ (JCPDS file, card no. 00-035-0762). As such, the 2$\theta$ range of 31$^0$ to 35$^0$ holds significant importance, as the XRD peaks of ZnO$_x$N$_y$, being an alloy of these two compounds, are expected to be situated between these two peaks.

Additionally, the hexagonal wurtzite structure of ZnO comprises four atoms per unit cell and belongs to the  $C^4_{6v}
(P6_3mc)$ space group\cite{wang2006raman}\cite{kaschner2002nitrogen}. It exhibits a particular Raman E2 phonon mode at 434 cm$^{-1}$\cite{yahia2008raman}. On the other hand, Zn$_3$N$_2$ is known to crystallize in the anti-bixbyite structure, which is characterized by a body-centered cubic lattice and belongs to the space group $Ia-3$\cite{nunez2012zinc}. A notable feature of pure Zn$_3$N$_2$ cubic lattice is the zero Raman shift observed in its spectrum\cite{khoshman2012growth}. 

Fig.\ref{fig3a} and \ref{fig3b} present the structural transition of ZnO$_x$N$_y$ thin films with band gaps through XRD and Raman patterns, respectively. The patterns have been differentiated through the use of various colours, each representing a specific band gap value. The yellow pattern corresponds to a band gap of 1.66 eV, while the purple pattern corresponds to 2.15 eV. The green pattern represents a band gap of 2.7 eV, and the grey pattern corresponds to 3.25 eV. 

In the case of ZnO$_x$N$_y$ thin films with the lowest band gap of 1.66 eV, XRD analysis reveals two prominent peaks at 32.89$^0$ and 32.98$^0$, accompanied by Raman peaks at shifts of 239, 352, and 589 cm$^{-1}$. The XRD peaks are attributed to the formation of an alloy within the film structure, blending the hexagonal wurtzite ZnO and cubic Zn$_3$N$_2$. The Raman peak at 239 cm$^{-1}$ is related to the Zn-N bonding configurations and is believed to result from local vibrational modes (LVMs) induced by intrinsic lattice defects in the Zn$_3$N$_2$ films\cite{addie2022effect}. The origin of the Raman band located in the range of 566 to 595 cm$^{-1}$ has been variously attributed by different researchers to local vibrational modes related to nitrogen, lattice defects within the host, and a fusion of local vibrational modes and disorder-activated scattering \cite{kerr2007raman}\cite{khoshman2012growth}\cite{bundesmann2003raman}. This peak has been observed in every sample studied. The peak at 353 cm$^{-1}$, on the other hand, has not been previously identified and is thought to be specific to the alloy structure. This peak decreases as the band gap increases and eventually disappears as the band gap reaches 3.3 eV, corresponding to pure ZnO.

As the band gap increases to 2.23 eV, the XRD patterns become broader, with peaks appearing at 32.77$^0$ and 32.84$^0$. And along with the previous peaks in the Raman spectrum, an additional peak at 445 cm$^{-1}$ has been detected, which is closely linked to the characteristic wurtzite peak of ZnO. This suggests that as the band gap reaches 2.23 eV, the previously established alloy structure begins to exhibit subtle alterations, leading to the emergence of a distorted ZnO structure.

At 2.78 eV, the XRD and Raman spectra indicate a transformation and progression of the film structure toward a more dominant ZnO character. The XRD peaks that correspond to the alloy phase exhibit a slight shift, whereas a striking characteristic peak of wurtzite ZnO at 34.08$^0$ is detected. Additionally, the Raman peak at 424 cm$^{-1}$ has shifted to 434 cm$^{-1}$, providing further validation of the formation of the ZnO phase. Consequently, in this band gap range, a co-existence of both the ZnO$_x$N$_y$ alloy phase and the wurtzite ZnO phase can be observed.

Finally, when the band gap exceeds 3 eV, the alloy phase completely vanishes, yielding a single XRD peak at 33.85$^0$ (JCPDS file, card no. 00-021-1486) of ZnO.  In the Raman spectra, the wurtzite peak of ZnO (434 cm$^{-1}$) becomes the dominant feature, while the peak at 350 cm$^{-1}$, previously related to the alloy structure, is no longer present. The 239 cm$^{-1}$ peak is still present with decreased intensity, implying that it is related to some intrinsic defects due to nitrogen. Additionally, the Raman spectra of the 3.3 eV sample reveal several peaks that are closely linked to the wurtzite structure. Of particular significance is the 275 cm$^{-1}$ peak, which is associated with N-doped ZnO and has been argued to represent a B$^{low}_1$ mode of disorder-activated Raman scattering, a mode that is forbidden in wurtzite ZnO\cite{kaschner2002nitrogen},\cite{kennedy2008raman},\cite{liu2011doping}. The 619 and 670 cm$^{-1}$ peaks, identified by Ribut et al.\cite{ribut2019investigations}, are related to ZnO deposited on Si substrates. Thus, the results suggest that a wurtzite ZnO structure with a high degree of structural disorder due to nitrogen doping has been formed when the band gap surpasses 3 eV.

\subsection{Morphological Evolution}
\label{subsection7}

\begin{figure*}
\centering
\subfigure{
   \includegraphics[width=8cm]{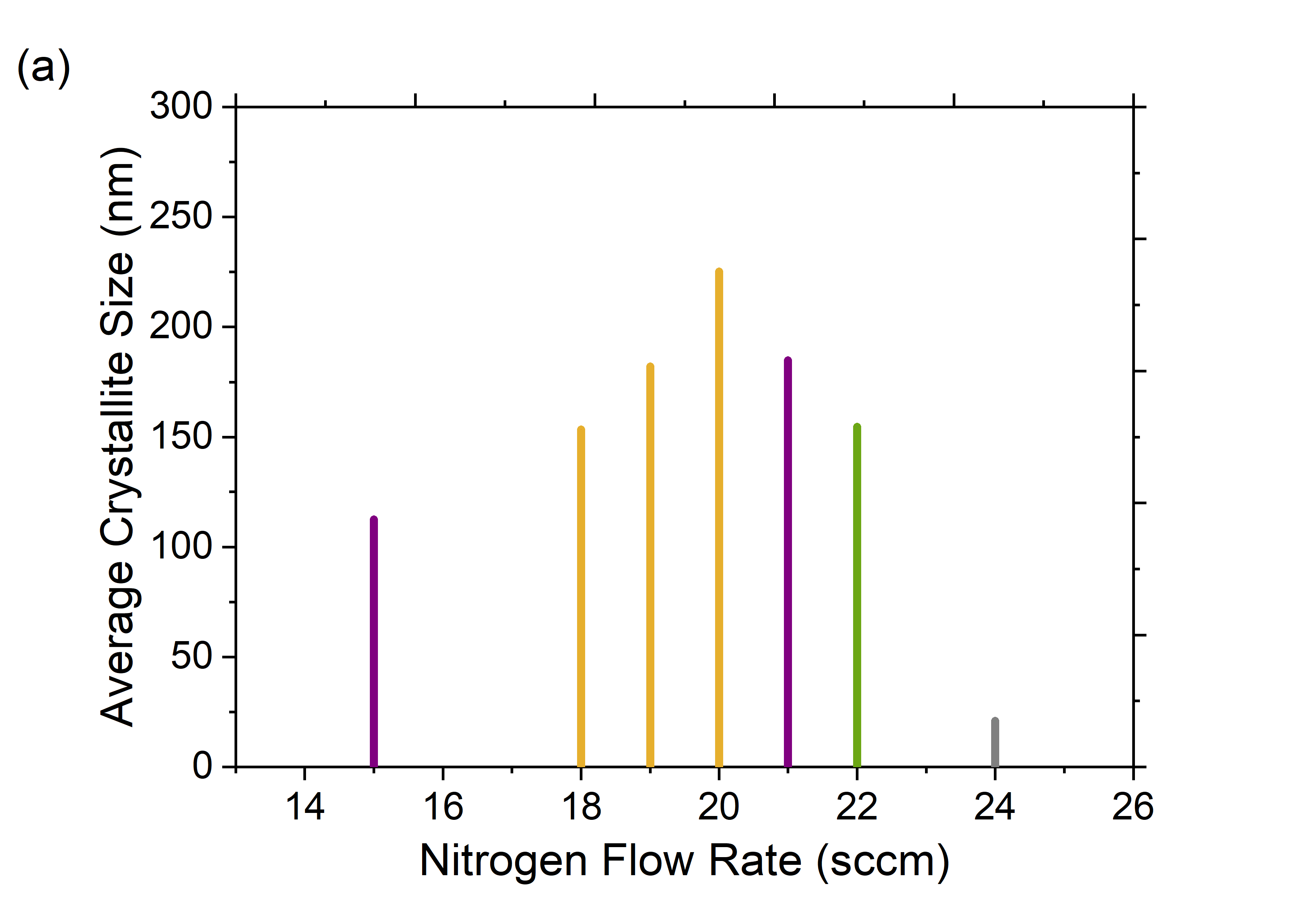}
   \label{fig4a}
 }
 \subfigure{
   \includegraphics[width=8cm]{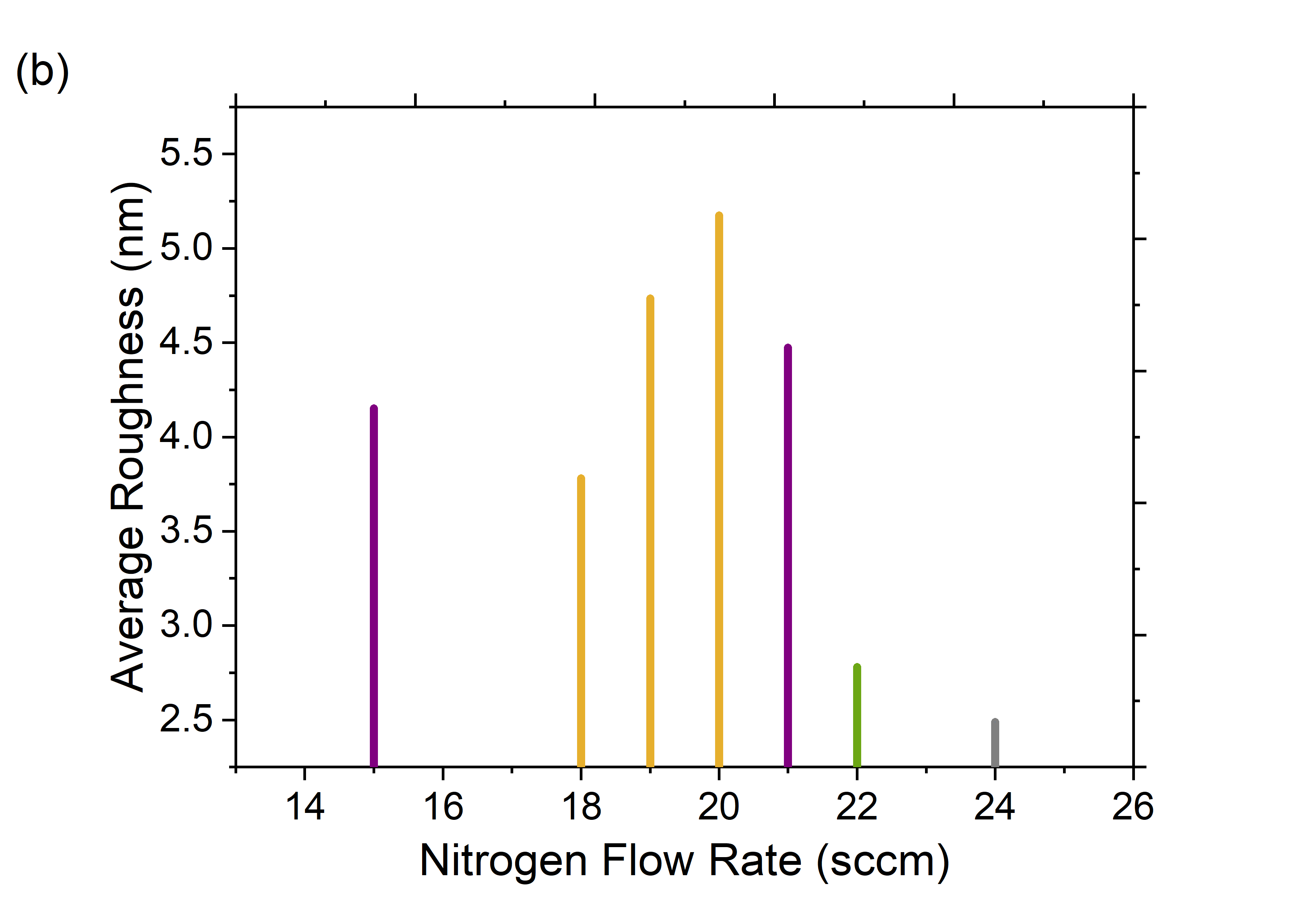}
   \label{fig4b}
 }

\label{4}
\caption{Variation of crystallite size and Roughness with Nitrogen Flow Rate. (a) Variation of crystallite size with respect to the nitrogen flow rate, (b) Variation of roughness with respect to the nitrogen flow rate.}
\end{figure*}

The morphological transition in the thin film was analyzed through the use of XRD and AFM techniques. To determine the average size of crystalline particles, the well-known Scherrer equation was employed in the analysis. The equation calculates crystallite size, represented by D, through a combination of various variables including the Scherrer constant (K), X-ray wavelength ($\lambda$), full width at half maximum (FWHM) of the XRD peak ($\beta$), and the Bragg angle ($\theta$), given by\cite{patterson1939scherrer},

\begin{equation} \label{e4}
\begin{array}{ll}
D = K\lambda/\beta cos(\theta)
\end{array}
\end{equation}

For samples presenting multiple peaks, a weighted average was determined using the intensity of each peak as a weighting factor. AFM allowed for the acquisition of high-resolution images of the thin film surface, enabling a comprehensive evaluation of its morphological features, including grain size, shape, and distribution, as well as roughness.

The effect of the nitrogen flow rate on crystallite size is presented in Fig.\ref{fig4b}. It was observed that the film with the largest crystallite size was formed at a flow rate of 20 sccm. As the flow rate deviated from this value, either by increasing or decreasing, the crystallite size decreased. As depicted in Figure 3a, the XRD peaks of the alloy phase were found to be very sharp. In the transition states, however, the peaks became broader and the crystallite size started to decrease. When the film reached the wurtzite phase, the crystallite size was significantly reduced due to the high working pressure\cite{shah2010microstructural}, which resulted in the formation of a disordered phase with numerous nitrogen-induced defects.

\begin{figure*}[t!]
\centering
   \includegraphics[width=15cm]{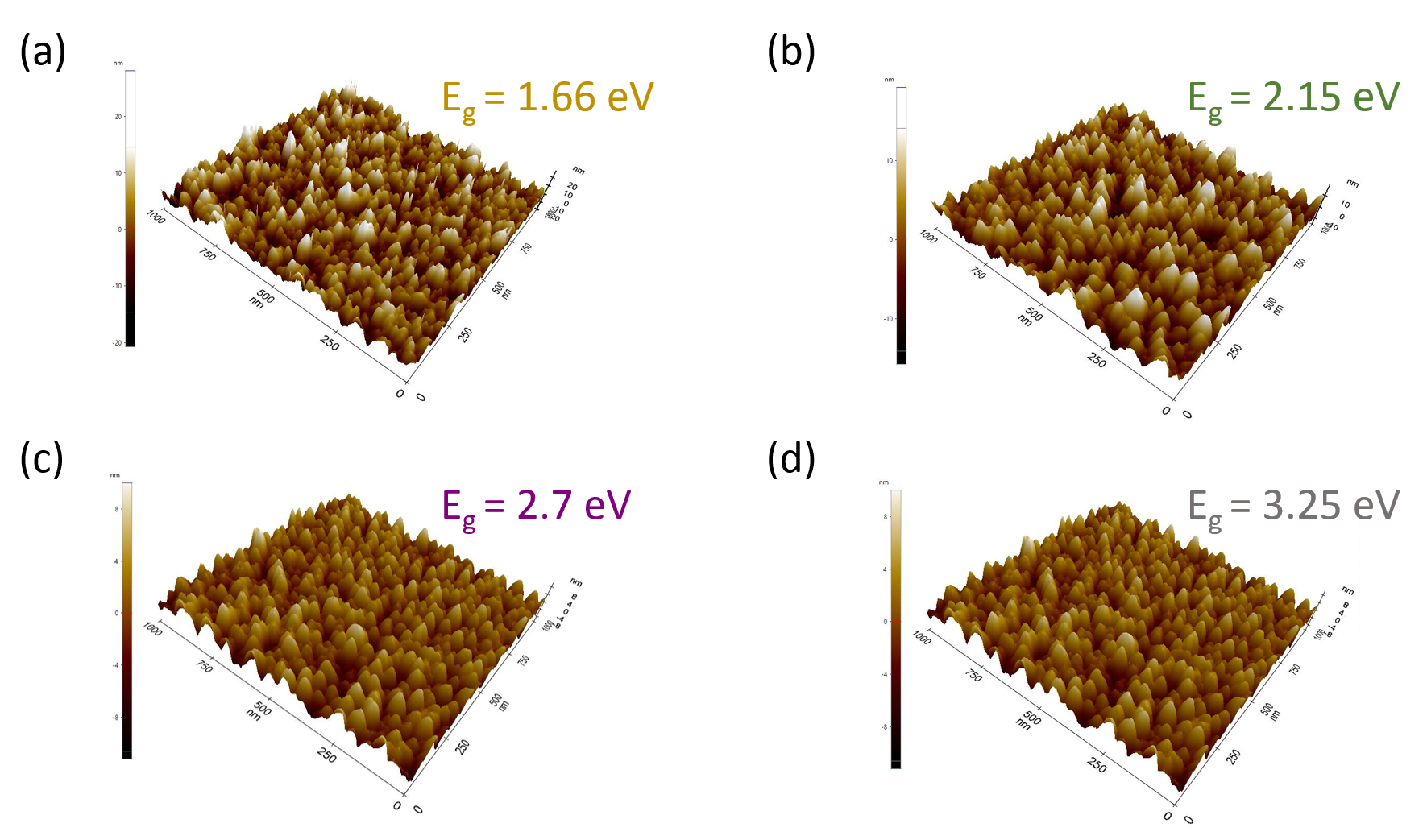}
 \label{fig5}
\caption{AFM Images of Thin Film at Different Band Gaps. (a) 1.66 eV, (b) 2.15 eV, (c) 2.7 eV, (d) 3.15 eV.}
\end{figure*}

The pattern of roughness in the thin film followed a similar trend as the crystallite size (Fig.\ref{fig4b}). Increased crystallinity of the film resulted in a corresponding increase in roughness, with the maximum roughness observed in the film formed at 20 sccm with the minimum band gap and optimal crystal quality. Films formed at flow rates greater than 20 sccm showed a sharp decrease in roughness, which was attributed to both the high working pressure\cite{devaraj2016fabrication} and the higher degree of amorphous behavior (smaller crystallite size). The variation in film roughness is further visualized through 3D images presented in Fig.\ref{fig5}. The AFM images showed that the film became smoother as the band gap increased, although no clear trend was observed with respect to the band gap. The roughness of the film was found to be influenced by both the working pressure and the structural transitions, with both factors contributing equally to the observed variation.

\subsection{Urbach Tail States}
\label{subsection8}

The Urbach tail is a term used to describe the tail of the density of states (DOS) at the bottom of the conduction band of semiconductors\cite{abe1981interband}\cite{chung2023visible}. The tail is due to the presence of impurities and defects in the material, which introduces additional energy states near the band gap\cite{moustafa2009growth}. In the case of (ZnO$_x$N$_y$) thin films, the Urbach tail can be attributed to the presence of oxygen vacancies and nitrogen interstitials\cite{dhara2012stable}\cite{zhou2013n}. These defects act as shallow acceptors and can introduce energy states within the band gap, which can affect the electronic properties of the material\cite{satapathy2018optical}.

 The Urbach energy is closely related to the band tail width of localized states in semiconductors, and an increase in Urbach energy can be attributed to an increase in defect states\cite{hassanien2015influence}. The presence of disorder and defects within thin films can result in the emergence of localized states close to the conduction band level, which causes an expansion of the band tail width ($E_U$). 

\begin{figure}[ht]
\centering
  \includegraphics[width=13cm]{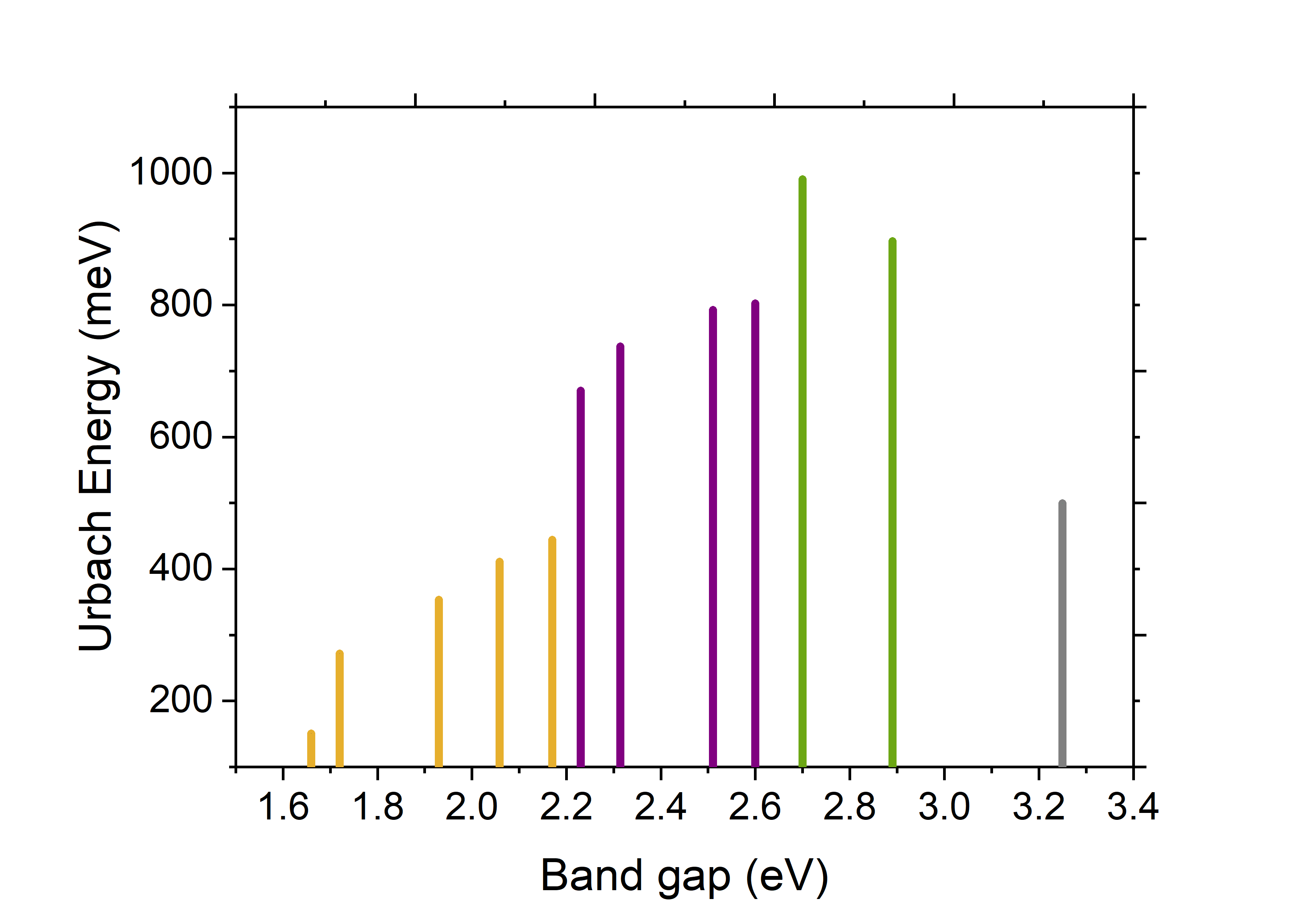}
  \caption{Figure 6: Variation of Urbach Energy with Band Gap.}
  \label{fig6}
\end{figure}

As depicted in Fig.~\ref{fig6}, the Urbach energy in ZnO$_x$N$_y$ thin films was found to have a correlation with the band gap. Despite the indirect influence of the nitrogen flow rate on the defect states, which in turn affects the crystallite size, the effect of the structural transition on the Urbach energy was more prominent. The energy initially rises with the increase in the band gap, reaching a peak before decreasing. This upward trend can be attributed to the transition from an alloy state to a mixture of alloy and wurtzite phases, resulting in an increase in structural disorders, dislocations, and vacancies. The peak of Urbach energy corresponds to the coexistence of both phases, where disorder is at its highest. The decline in Urbach energy can be explained by the domination of the wurtzite structure, although the energy remains significantly higher compared to that of ZnO thin films (80-90 meV)\cite{archana2022influence}, indicating the presence of a distorted lattice. It is clear that the evolution of tail states is inextricably linked to structural evolution and follows a similar pattern.

\section{Conclusion}
\label{section4}
In conclusion, a novel strategy for tuning the band gap of ZnO$_x$N$_y$ films has been developed through the manipulation of the working pressure to adjust the mean free path of nitrogen ions during sputtering. The structural evolution of the film with changing band gap was extensively studied, and intermediate structures were identified through XRD and Raman spectroscopy. The initial alloy structure of the film was found to exist from 1.66 eV to 2.15 eV, beyond which a distorted wurtzite structure began to emerge, as indicated by the 451 cm$^{-1}$ peak in the Raman spectra. At a band gap of 2.74 eV, the peak shifted to 434 cm$^{-1}$, becoming more prominent and indicating the coexistence of both alloy and wurtzite structures. With an increasing band gap, the wurtzite structure became dominant, completely replacing the alloy structure at 3.25 eV. Additionally, a special Raman peak at 350 cm$^{-1}$ was identified and thought to be associated with the alloy phase, which disappeared in the fully developed wurtzite phase. The study of the Urbach tails also revealed that the disorder in the film was maximum when two structures coexisted. These findings offer a profound understanding of the structural evolution of ZnO$_x$N$_y$ films and pave the way for further advancements in optoelectronic devices.

\section{Acknowledgement}

The authors would like to express their gratitude to the various organizations and individuals who supported and contributed to this research. The Department of Science and Technology, SERB provided a Start-Up Grant from 2021 to 2023 and a Faculty Research Grant (FRG) at the National Institute of Technology Calicut (NITC) provided financial support. The Center for Materials Characterization (CMC) at NITC provided access to the Raman and X-ray Diffraction (XRD) facilities and the authors would like to thank Mr. Shintu Varghese for his technical assistance with the Raman facility, Mr. Nithin for his technical support with the XRD analysis, and the staff at CeNS at the Indian Institute of Science (IISc) for their technical support with the AFM facility. The authors also appreciate the assistance of the Department of Chemistry and Mr. Diljith with the UV-Visible Spectrophotometer.

\bibliographystyle{unsrt}  
\bibliography{references}

\end{document}